\documentclass[twocolumn,trackchanges,twocolappendix]{aastex701}

\usepackage{amsmath}
\begin{document}

\title{Coupled orbital and interior structure evolution of lava planets}

\author[0009-0004-3980-8143]{Mahesh Herath}
\affiliation{Department of Earth \& Planetary Science, McGill University, 3450 Rue University, Montreal, QC, Canada, H3A 0E8}
\affiliation{Trottier Space Institute, McGill University, 3550 Rue University, Montreal, QC, Canada, H3A 2A7}
\email{mahesh.herath@mail.mcgill.ca} 

\author[0000-0001-6129-5699]{Nicolas B. Cowan}
\affiliation{Department of Earth \& Planetary Science, McGill University, 3450 Rue University, Montreal, QC, Canada, H3A 0E8}
\affiliation{Trottier Space Institute, McGill University, 3550 Rue University, Montreal, QC, Canada, H3A 2A7}
\affiliation{Department of Physics, McGill University, 3600 Rue University, Montreal, QC, Canada}
\email{nicolas.cowan@mcgill.ca}

\author[0000-0002-9265-4209]{Charles-\'Edouard Boukar\'e}
\affiliation{Department of Physics \& Astronomy, York University, 140 Campus Walk Room 128, North York, Toronto, ON, Canada}
\email{boukare@yorku.ca}

\author[0000-0001-5677-1582]{Mathieu Dumberry}
\affiliation{Department of Physics, University of Alberta, 11335 Saskatchewan Drive NW, Edmonton
AB, T6G 2H5, Canada}
\email{dumberry@ualberta.ca}

\begin{abstract}

Lava planets 
likely did not form in their current orbits, instead migrating inward via orbital decay, which influenced the evolution of their magma oceans. We introduce a coupled thermal-orbital evolution model to explore how rocky planets migrate from the inner edge of the protoplanetary disk, with periods of 1--10 days, to orbital periods of less than a day. In our model, mantle melting is controlled by tidal heating and stellar flux, while orbits evolve via tidal migration. The mantle’s tidal quality factor varies with its temperature and structure, creating a feedback loop between thermal evolution and orbital decay. We use our numerical model to simulate the migration of seven known lava planets: K2-141b, K2-360b, TOI-141b, TOI-431b, TOI-2431b, HD\,3167b and GJ\,367b. Migration occurs in two stages: an initial high-eccentricity stage reducing the semi-major axis by a factor of $\sim 2$, followed by a low-eccentricity stage reducing it by a factor of $\sim 5$. A successful migration from $\sim 0.1$ AU to a present-day orbit requires starting eccentricities $\ge 0.9$ and sustained eccentricity forcing with $e_{\mathrm{min}} \ge 10^{-2}$. The rate of migration depends on the state of the mantle: slow when mostly molten, fast when mostly solid. This pathway works for most lava planets, but not for TOI-431b or GJ-367b, suggesting that multiple migration pathways are possible for lava planets.

\end{abstract}



\section{Introduction} \label{sec1}

Lava planets are rocky exoplanets that are tidally locked and orbiting so close to their host star that one side is molten while the other is solid \citep{Schaf2009, Scheider2011, Kite2016, Lichtenberg2021}. The stellar irradiation at their current orbits is high enough to evaporate the material that rocky planets are made of; therefore, it is unlikely that lava planets formed in-situ \citep{Ogihara2015}. They must have formed further out before migrating inwards \citep{Lee2017}. 

Some proposed migration mechanisms for lava planets are high eccentricity migration and low eccentricity secular scattering \citep{Petro2018, Pu2019}. Low-eccentricity secular scattering occurs when planets spaced far apart have non-periodic interactions between them, leading to small eccentricity boosts. High-eccentricity migration can be caused by mean-motion resonances or the Kozai-Lidov mechanism, leading to large eccentricity boosts. In \citet{Leconte2010}, \citet{Pu2019}, and \citet{Schmidt2024} it was shown that orbital circularization occurs early in the evolution of rocky planets, and external excitation of eccentricity is required for them to become ultra-short-period planets (USP's). \citet{Petro2018} demonstrated that planets orbiting between 0.05 and 0.1 AU can reach high eccentricities ($0.5 < e < 0.9$) via secular scattering and migrate to within two stellar radii before tidal capture. 

However, previous authors have not considered the feedback between orbital and structural evolution. The rate of migration depends on tidal dissipation within the mantle of a planet.  Tidal dissipation, in turn, depends on the mantle viscosity, and thus it depends on the material state of the mantle.  That is, it depends on how much of the mantle is solid, mush, or liquid.  Yet, the state of the mantle depends on its interior temperature profile, which is altered by tidal heating.  As a planet migrates inward, its mantle state therefore evolves in response to tidal heating, and this evolution has a feedback on tidal dissipation. 
Inclusion of this feedback loop is important to properly determine how long the mantle remains in a molten mantle, the timing of subsequent events of mantle melting, and the timescale of  migration of lava planets toward their present-day orbits. 

In this paper, we present a coupled orbital-thermal model to investigate how lava planets may migrate to their present-day, short-period orbits. We explore potential correlations between migration history and current thermal state. In this model, tidal heating is directly linked to the mantle structure as well as the orbital eccentricity and semi-major axis. We apply this model to seven lava planets: K2-141b, K2-360b, TOI-141b, TOI-431b, TOI-2431b, HD-3167b and GJ-367b. These planets evolve over 10 Gyrs until reaching their present-day orbits. We start by describing the model (Section \ref{sec2}), followed by a presentation of the results (Section \ref{sec3}). We finish the paper with a discussion (Section \ref{sec4}) on what our findings imply about the formation of lava planets.

\section{Model} \label{sec2}

\subsection{Mantle viscosity and the tidal dissipation factor}

The viscoelastic response of the planetary interior to tidal forces is determined by the mechanical properties of the interior layers, in particular their rigidity and viscosity. As described in \citet{Herath2024}, the mantle of a lava planet is composed of layers of liquid (magma), partial melt (mush), and solid rock. A liquid metallic iron core is assumed to occupy the central (lower-most) region of the planet (see Figure \ref{fig1}), which is assumed to have zero rigidity and a low viscosity that does not change with time \citep{Nichols2024}. 
We model the viscosity of each of the mantle layers based on the melt fraction ($\phi$) most likely for their phases of matter; the liquid, mush and solid layers of the mantle are assumed to have melt fractions close to 100 percent, 35 percent and 1 percent, respectively. For simplicity, we keep these melt fractions fixed within each layer. To calculate the resulting viscosities ($\eta$) from these melt fractions, we use the equations given in \citet{costa2009} and \citet{Kervazo2021}:

\begin{equation}
\label{Eqn1}
\eta (\phi) = \frac{ \left(1 + \Theta^{\delta} \right)}{\left[1 - F\left(\Theta, \xi, \gamma \right)\right]^{B\phi_{*}}},
\end{equation}

\noindent where $B=2.5$ is the Einstein coefficient  \citep{costa2009}, $\phi_{*}=0.569$ is the critical melt fraction, $\delta=25.7$ \citep{Kervazo2021}, and $\Theta$ is the normalized melt fraction given as

\begin{equation}
\label{Eqn2}
\Theta = \frac{\phi}{\phi_{*}}.
\end{equation}

\noindent The melt fraction $\phi$ takes a value between 0 and 1, depending on the percentage of melt in a given layer. The auxiliary function $F(\Theta, \xi, \gamma)$ is evaluated as

\begin{equation}
\label{Eqn3}
F = \left( 1 - \xi \right) \mathrm{erf}\left[ \frac{\sqrt{\pi}}{2\left( 1 - \xi\right)} \Theta \left( 1 + \Theta^{\gamma}\right) \right], 
\end{equation}

\noindent where our adopted choices for $\xi$ and $\gamma$ are listed in Table \ref{tab1}. The parameter $\gamma$ measures the rapidity at which the relative viscosity increases with crystal fraction. With the melt fractions fixed, the viscosities of the liquid, mush and solid layers are approximately 1, $10^{8}$, and $10^{19}$ Pa s respectively.

The internal heat generated by tidal forcing depends on the tidal quality factor $Q$, which gives the ratio of maximum potential energy stored in a planet's tidal bulge during an orbital period, to the energy released from the bulge per tidal cycle. The tidal quality factor is tied to the viscosity and mass of the mantle layers and is therefore directly affected by the internal structure \citep{Tobie2019,organowski20,Kervazo2021}. Our model for $Q$ is rooted in an analytical model of tidal dissipation for a uniform planet. We use the tidal love number $k_{2}$, which represents the added gravitational potential at the surface of a planet caused by tidal deformation \citep{Seager2010}. The tidal Love number for a uniform planet is:

\begin{equation}
\label{Eqn4}
k_{2} = \frac{1.5}{\left(1 + \frac{19}{2}.\frac{\mu}{\rho g R_{p}}\right)},
\end{equation}

\noindent where $\rho$ is the density, $g$ is the gravitational acceleration at the surface, $R_p$ the planetary radius, and $\mu$ the shear modulus.  The latter is determined based on an Andrade rheological model \citep{Castillo2011, Henning2014, Tobie2019}, 

\begin{equation}
\label{Eqn5}
\mu = \frac{\mu_{r}}{J},
\end{equation}

\noindent where $\mu_{r}$ is the shear modulus in the pure elastic limit and $J$ a complex compliance given by 

\begin{equation}
\label{Eqn6}
J = 1 - \frac{i}{\chi} + \frac{\Gamma \left(1 + \alpha \right)}{i \chi^{\alpha}},
\end{equation}

\noindent where $\Gamma$ is the gamma function and $\alpha$ describes the frequency dependence of the transient response; we assume $\alpha = 0.3$ \citep{Tobie2019}. The tidal frequency factor $\chi$ is related to the Maxwell relaxation time $\tau_{m}=\frac{\eta}{\mu_{r}}$ and the orbital frequency $\omega$ by 

\begin{equation}
\label{Eqn7}
\chi = \omega \tau_{m}\, .
\end{equation}

\noindent The orbital frequency is dependent on the orbital radius, and therefore changes as the planet migrates. Finally, $Q$ is obtained from $k_{2}$ and its imaginary component:

\begin{equation}
\label{Eqn8}
Q = \left|\frac{k_{2}}{- \mathrm{Im} \left[k_{2}\right]}\right|.
\end{equation}

\noindent For a given viscosity $\eta$, when the orbital period $2\pi/\omega$ is much shorter than the Maxwell relaxation time $\tau_{m}$, then $\chi \gg1$, $J\approx1$ and the rigidity of the planet approaches that in its elastic limit, $\mu \approx \mu_r$.  This is the situation for the diurnal tides on Earth, and the observed $k_{2}\approx0.3$ \citep{Ray2001} implies a mean elastic rigidity of $\mu_{r}=1.5 \times 10^{11}$ Pa, consistent with PREM \citep{prem}; we assume the same value for the lava planets in our study.  

To obtain $Q$ for a layered mantle structure, we proceed as follows.  We compute the $Q$ values at orbital frequency $\omega$ for uniform planets of radius $R_p$ and mass $M_p$ for the viscosities of the liquid, mush and solid layers computed above.  This gives us individual values of $Q$ for a liquid planet ($Q_{l}$), mush planet ($Q_{m}$) and solid planet ($Q_{s}$).  We then compute the planetary bulk dissipation factor $Q_{p}$ based on an average weighted by the individual masses of liquid ($M_{l}$), mush ($M_{m}$), and solid ($M_{s}$) layers:

\begin{equation}
\label{Eqn9}
\frac{M_{l}}{Q_{l}} + \frac{M_{m}}{Q_{m}} + \frac{M_{s}}{Q_{s}} = \frac{M_{p}}{Q_{p}} \, . 
\end{equation}

\noindent This harmonic mean ensures that the regions of the mantle with the lowest $Q$, the mush layer, dominates the global $Q$. A more rigorous way to calculate $Q_p$ for a layered planet involves integrating in radius the set of elasto-gravitational equations subject to a periodic tidal forcing \citep{organowski20}. However, given the number of assumptions and approximations that enter our mantle model (uniform layers, fixed melt fractions, etc.), a more precise calculation of $Q_p$ for such a crude mantle model would still remain, at best, an order of magnitude estimate. In this context, our simple model, although crude, is sufficient to capture, in an order of magnitude sense, how $Q_p$ depends on the interior state of the mantle. 

To provide an example of how $Q_{p}$ varies with mantle structure and different mixtures of silicate phases, we consider a lava planet with a radius of 7000 km, with a mantle thickness of 3400 km and a core radius of 3600 km, orbiting at a frequency of 6.7 hours, equivalent to that of K2-141b \citep{Barragan2018}. Figure \ref{fig2} (solid line) shows how $Q_p$ varies with $\eta$ based on Equations \ref{Eqn4} to \ref{Eqn8} assuming a uniform bulk mantle viscosity. This demonstrates how $Q_{p}$ changes with viscosity under an Andrade rheology model. We also show on the same figure specific values of $Q_{p}$ computed for four different mantle states. The mixtures in terms of volume are: 80 percent mush and 20 percent liquid, 20 percent liquid, 40 percent mush and 40 percent solid, 50 percent mush and 50 percent solid, 20 percent mush and 80 percent solid. We use the Arrhenius logarithmic mixing rule and Krieger-Dougherty equations to estimate bulk mantle viscosities for these mixtures. The $Q_{l}$, $Q_{m}$ and $Q_{s}$ values for each of these mixtures are then computed from Equations \ref{Eqn4} to \ref{Eqn8} using viscosities of 1, $10^{8}$ and $10^{19}$ Pa s for liquid, mush and solid respectively, and $Q_{p}$ is then computed from Equation \ref{Eqn9}. As shown in Figure \ref{fig2}, the resulting $Q_{p}$ for each mantle mixture matches well the predictions based on a uniform mantle of equivalent bulk viscosity.


\subsection{Thermal evolution}

We assume a rocky planet with an Earth-like Silicate mantle composition and an Earth-like core mass fraction, consistent with the measured bulk densities of small exoplanets \citep{Brinkman2024}. The bulk density of the mantle ($\rho_{m}$) is assumed uniform and equal to 4600 kg m$^{-3}$. Other thermoelastic constants used for the mantle are listed in Table \ref{tab1}. We assume the mantle is convecting at all times and we adopt an adiabatic temperature profile (geotherm) within each layer of the mantle. We follow \citet{Lab2003,Lab2015}, and connect the temperature at the top of a layer ($T_{top}$ at radius $r_{\mathrm{top}}$) to that at the bottom ($T_{\mathrm{bottom}}$, $r_{\mathrm{bottom}}$) with the relation

\begin{equation}
\label{Eqn10}
T_{\mathrm{bottom}} = T_{\mathrm{top}} \exp \left(\frac{r_{\mathrm{top}} - r_{\mathrm{bottom}}}{D }\right) \, .
\end{equation}

\begin{figure}\centering
\includegraphics[width=1.0\columnwidth]{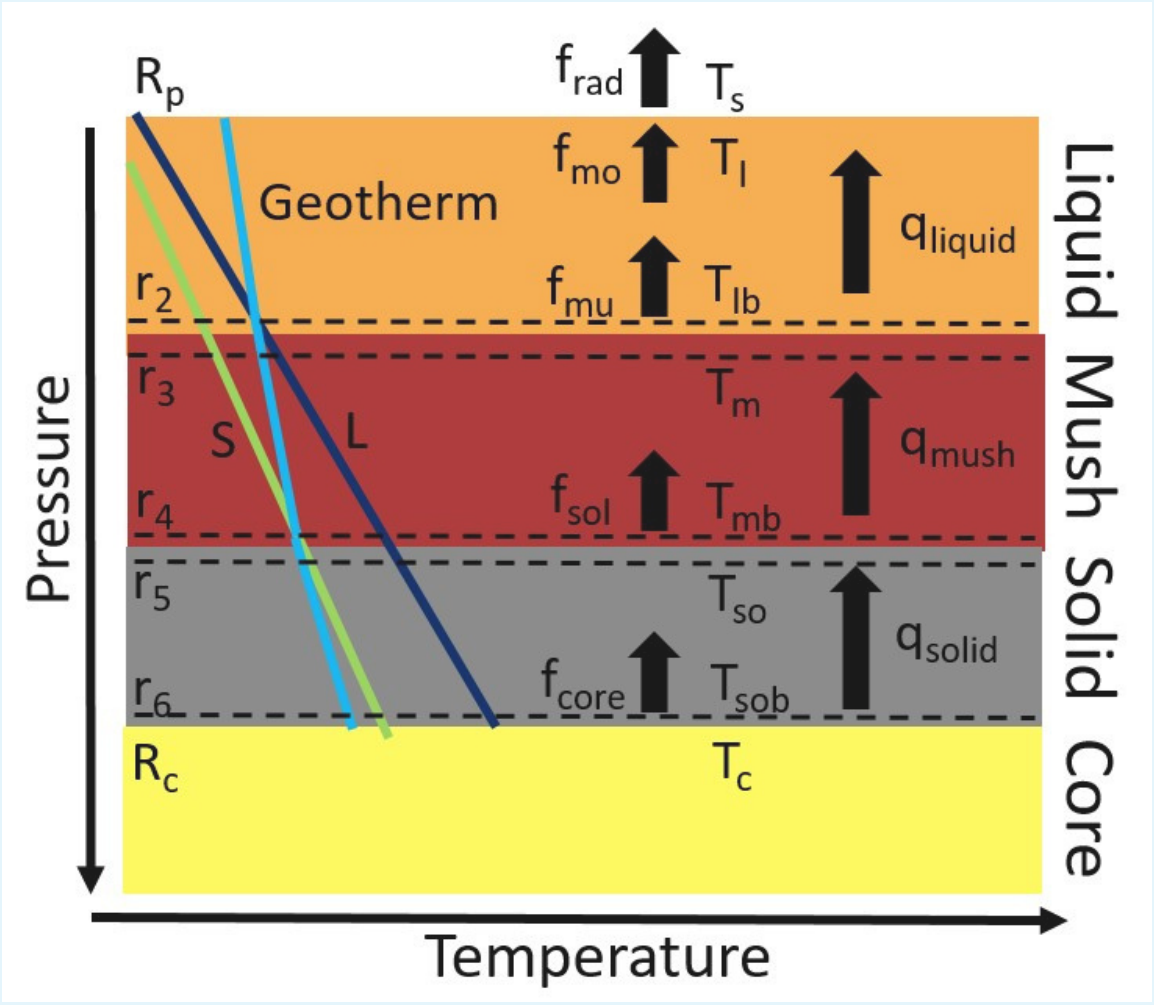}

\caption{Schematic of the 1D cross section and thermal evolution model from \citet{Herath2024}. $f_{\textrm{rad}}$ is the radiated flux, $f_\textrm{mo}$ the heat flux from the magma ocean, $f_\textrm{mu}$ the heat flux from the magma mush, $f_\textrm{sol}$ is the flux from solid mantle, and $f_\textrm{core}$ the flux from the core. $q_\textrm{liquid}$, $q_\textrm{mush}$ and $q_\textrm{solid}$ are the tidal dissipation in the liquid, mush and solid, respectively. $T_{l}$, $T_{m}$, and $T_{so}$ denote the temperatures at the tops of the liquid, mush, and solid layers respectively. $T_{lb}$, $T_{mb}$, and $T_{sob}$ denote the temperatures at the bottoms of each layer, in the same order. $r_{2}$, $r_{3}$, $r_{4}$, $r_{5}$ and $r_{6}$ denote the radii of the thermal boundaries at the intersections between layers}. $T_{s}$ is the surface temperature, while $T_{c}$ is the core temperature. The light blue line is the geotherm of the mantle while the line ‘L’ (dark blue line) is the mantle liquidus and ‘S’ (green line) is the mantle solidus. The pressure increases downward while the temperature increases to the right.
\label{fig1}
\end{figure}

\begin{figure}\centering
\includegraphics[width=1.0\columnwidth]{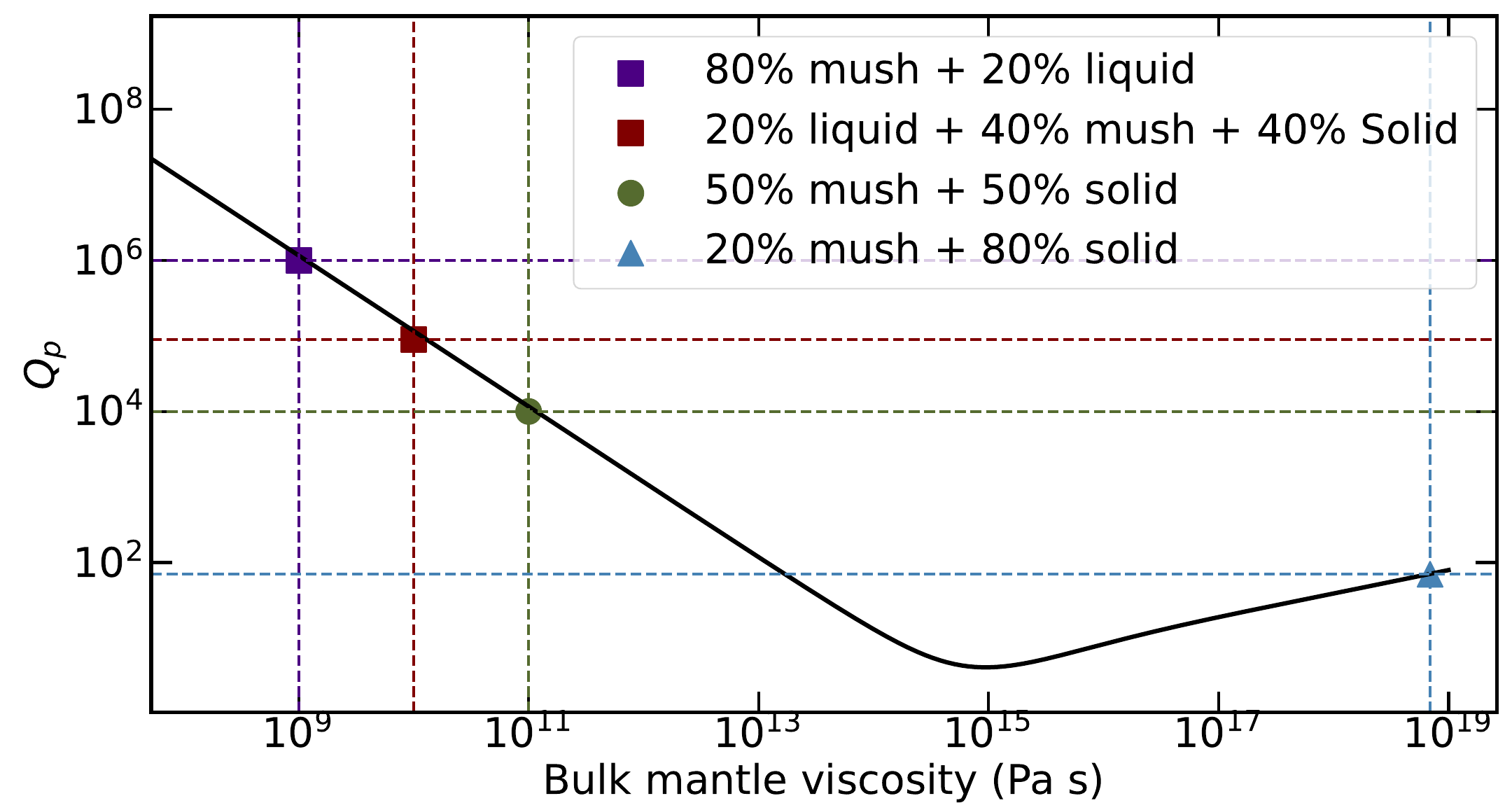}

\caption{The tidal quality factor $Q_{p}$ as a function of bulk mantle viscosity based on an Andrade rheology model for the lava planet K2-141b at an orbital frequency of 6.7 hours and a mantle with a thickness of approximately 
$0.6R_{\oplus}$. The solid line shows how $Q_p$ varies with viscosity for a uniform molten mantle. We did not have any simulated data points to include viscosities below $10^{7}$ Pa s wherein the viscosity at that point would be approximately 1 Pa s. It should be noted that this total molten state is brief because heat loss is rapid and would result in mush and mush crystals in less than 1000 years (as seen in Figure 6, panels k and l ). Symbols (see legend) show how $Q_p$ varies for specific mantle mixtures (see text for details).}
\label{fig2}
\end{figure}

We apply this relation to each of the liquid, mush and solid layers shown in Figure \ref{fig1}, with its appropriate values substituted for $T_{\mathrm{top}}$, $r_{\mathrm{top}}$ and $T_{\mathrm{bottom}}$, $r_{\mathrm{bottom}}$, which evolve with time. The factor $D$ is the temperature scale height which is evaluated by  

\begin{equation}
\label{Eqn11}
D = \sqrt{\frac{3C}{2 \pi \alpha \rho G}},
\end{equation}

\noindent where $C$ is the specific heat capacity and $\alpha$ is thermal expansivity.  We use the same values of $C$, $\alpha$ and $\rho$ in each of the mantle layers (given in Table \ref{tab1}), so $D$ is identical in each layer. Our initial conditions are: a fully molten mantle with an imposed temperature of 4000 K at $T_{s}$, $T_{lb}$ computed with Equation \ref{Eqn10} using the initial value of $T_{s}$, and an imposed temperature of $T_{c}=10^{4}$ at the core-mantle boundary. The temperature profile in the core may be specified as a parabolic function in radius with no heat flux at the centre, $T = T_o - c(r/R_c)^2$, with the central temperature $T_o$ and $c$ specified such that the heat flux matches that of the bottom boundary layer of the mantle at $r=R_c$, $T=T_c$. However, other than $T_c$, the core temperature profile does not influence the evolution of the mantle temperature profile. 

We track the evolving location and temperature of the boundaries between each layer as the planet migrates. The planetary surface temperature $T_{s}$ is allowed to vary with the equilibrium temperature $T_{eq}$ resulting from the incident stellar flux. The boundaries between liquid, mush and solid states are determined with the liquidus ($T_{\mathrm{liq}}$) and solidus ($T_{\mathrm{sol}}$) curves from \cite{Fiq2010} and \cite{zhang1994}: 

\begin{equation}
\label{Eqn12}
T_{\mathrm{liq}} = 2000 \left(0.1169 \left(\frac{P}{\mathrm{GPa}}\right) + 1\right)^{0.32726}
\end{equation}

\begin{equation}
\label{Eqn13}
T_{\mathrm{sol}} = 1674 \left(0.0971 \left(\frac{P}{\mathrm{GPa}}\right) + 1\right)^{0.35175},
\end{equation}

\noindent where $P$ is the hydrostatic pressure. The intersection of the geotherm with $T_{\mathrm{liq}}$ and $T_{\mathrm{sol}}$ specify the radii of the liquid-to-mush and mush-to-solid transitions, respectively. 

To track the heat flowing between the core and the mantle layers, we use a single-column version of the model described in \citet{Herath2024}. The state of the mantle changes between liquid, mush, and solid, with the evolution of the internal energy budget. The mantle energy budget is determined by the balance between the energy radiated from the surface ($f_{\textrm{rad}}$), the tidal dissipation within the layers of the mantle and the upward convective heat transport from the layers below. Latent heat has a negligible effect on these timescales \citep{Herath2024}. We also neglect radiogenic heating; as shown in Figure 6 of \citet{Herath2024}, tidal heating in lava planets is many orders of magnitude larger than the expected tens of TW of radiogenic heat flow of a planet with an Earth-like composition \citep[e.g.][]{lay08}. 

The net radiated power is computed from the Stefan-Boltzmann law:

\begin{equation}
\label{Eqn14}
f_{\mathrm{rad}} =  4 \pi R_{p}^{2}\sigma \left(T_{\mathrm{s}}^{4} - T_{\mathrm{eq}}^{4}\right),
\end{equation}

\noindent where the equilibrium temperature is

\begin{equation}
\label{Eqn15}
T_{\mathrm{eq}} =  T_{\mathrm{eff}}  \left(1 - A_{B} \right)^{\frac{1}{4}} \sqrt{\frac{R_{s}}{2a}}.
\end{equation}

\noindent Here, $T_{\mathrm{eff}}$ is the stellar effective temperature, $A_{B}$ is the planet's Bond albedo (which we assume to be zero), $a$ is the semi-major axis which changes throughout the simulation due to orbital migration, and $R_s$ is the stellar radius. The model of \citet{Herath2024} includes day–night heat transfer, but we neglect horizontal transport since the planet cannot be synchronously rotating for most of its migration due to significant eccentricity. Interior thermal evolution is controlled by heat loss on the night side, the only region where interior heat can escape. When the magma ocean is shallow, \citet{Kite2016} showed that day–night transfer is insufficient to warm the night side. When the magma ocean is fully molten from a hot core, \citet{Boukare2025b} (Section 5 of SI and Figure 7 of SI) showed that horizontal transfer is negligible relative to vertical transport. During migration, and at their present-day orbits, the magma oceans remain shallow with minimal internal differences between sides. 

The tidal dissipation in the mantle of a planet $q_{p}$ is related to the tidal quality factor $Q_p$ by \citep{Jackson2008}:  

\begin{equation}
\label{Eqn16}
q_{p} = \frac{e^{2}}{a^{\frac{15}{2}}}  \frac{63}{4}  \frac{G^{\frac{3}{2}}}{Q_{p}} M_{s}^{\frac{5}{2}} R_{p}^{5} \, ,
\end{equation}

\noindent where $G$ is the gravitational constant, $e$ is the orbital eccentricity, and $M_{s}$ is the stellar mass. This equation provides a measure of the tidal dissipation for the whole planet, which is used for orbital evolution. 
However, our model for the interior thermal evolution of a layered mantle requires expressions of tidal dissipation for each individual layers. To do so, we build expressions similar to Equation \ref{Eqn16} for each mantle layer, in which we replace $R_{p}$ with the outer radii of the layer in consideration at a given time step; that is, $R_{\mathrm{mush}}$ for the mush layer and $R_{\mathrm{solid}}$ for the solid layer (the outer radius of the liquid layer always remain equal to $R_{p}$).  Likewise, we replace $Q_p$ with its equivalent value for the liquid, mush and solid layers (that is, $Q_{l}$, $Q_{m}$, and $Q_{s}$). Our expressions for tidal dissipation in the liquid ($q_{\mathrm{liquid}}$), mush ($q_{\mathrm{mush}}$), and solid ($q_{\mathrm{solid}}$) layers are calculated as: 

\begin{equation}
\label{Eqn17}
q_{\mathrm{liquid}} = \frac{e^{2}}{a^{\frac{15}{2}}}  \frac{63}{4}  \frac{G^{\frac{3}{2}}}{Q_{l}} M_{s}^{\frac{5}{2}} R_{p}^{5} \, ,
\end{equation}

\begin{equation}
\label{Eqn18}
q_{\mathrm{mush}} = \frac{e^{2}}{a^{\frac{15}{2}}}  \frac{63}{4}  \frac{G^{\frac{3}{2}}}{Q_{m}} M_{s}^{\frac{5}{2}} R_{\mathrm{mush}}^{5} \, ,
\end{equation}

\begin{equation}
\label{Eqn19}
q_{\mathrm{solid}} = \frac{e^{2}}{a^{\frac{15}{2}}}  \frac{63}{4}  \frac{G^{\frac{3}{2}}}{Q_{s}} M_{s}^{\frac{5}{2}} R_{\mathrm{solid}}^{5} \, .
\end{equation}

To track the changing temperatures at the surface ($T_{s}$) and tops of the liquid layer ($T_{l}$), mush layer ($T_{m}$), and solid layer ($T_{so}$), the energy balance equations are expressed as

\begin{equation}
\label{Eqn20}
V_{\mathrm{mo}} C \rho_{m} \frac{dT_{s}}{dt} = q_{\mathrm{liquid}} + f_{\mathrm{mo}} - f_{\mathrm{rad}},
\end{equation}

\begin{equation}
\label{Eqn21}
V_{\mathrm{mo}} C \rho_{m} \frac{dT_{l}}{dt} = q_{\mathrm{liquid}} + f_{\mathrm{mu}} - f_{\mathrm{mo}},
\end{equation}

\begin{equation}
\label{Eqn22}
V_{\mathrm{mu}} C \rho_{m} \frac{dT_{m}}{dt} = q_{\mathrm{mush}} + f_{\mathrm{sol}} - f_{\mathrm{mu}},
\end{equation}

\begin{equation}
\label{Eqn23}
V_{\mathrm{sol}} C \rho_{m} \frac{dT_{so}}{dt} = q_{\mathrm{solid}} + f_{\mathrm{core}} - f_{\mathrm{sol}},
\end{equation}

\begin{equation}
\label{Eqn24}
V_{\mathrm{core}} C_{\mathrm{core}} \rho_{m} \frac{dT_{c}}{dt} = -f_{\mathrm{core}},
\end{equation}

\noindent where $f_{\mathrm{mo}}$, $f_{\mathrm{mu}}$, and $f_{\mathrm{sol}}$ are upward heat fluxes from the liquid, mush, and solid layers respectively. The term $f_{\mathrm{core}}$ is the heat flux from the core. The terms $V_{\mathrm{mo}}$, $V_{\mathrm{mu}}$, $V_{\mathrm{sol}}$, and $V_{\mathrm{core}}$ are the volumes of liquid, mush, solid, and core respectively. Terms $C$ and $C_{core}$ are the specific heat capacities of the mantle and core (see Table \ref{tab1}). It should be noted that the upward heat fluxes ($f$ variables) given in Equations \ref{Eqn20} to \ref{Eqn24} are integrated over the surface area of their respective layers (and are therefore in units of Watts). 

The process for computing the heat fluxes between each layer is based on Equations 12 to 22 in \citet{Herath2024} and can be summarized as follows.
The heat flux at a boundary is proportional to the temperature drop across this boundary and inversely proportional to the thickness of the temperature boundary layer. We then express the thickness in terms of the Rayleigh number and thickness of the layer below this boundary.  In this way, the heat flux leaving the magma ocean (liquid layer), $f_{\mathrm{mo}}$, depends on the difference between $T_{l}$ and $T_{s}$, the Rayleigh number of the liquid magma, and the liquid layer thickness (see Figure \ref{fig1}). Likewise, the heat flux leaving the mush, $f_{\mathrm{mu}}$ depends on the difference between $T_{m}$ and $T_{lb}$ and the Rayleigh number and thickness of the mush layer and its thickness, and the heat flux leaving the solid layer, $f_{\mathrm{sol}}$, depends on the difference between $T_{so}$ and $T_{mb}$ and the Rayleigh number and thickness of the solid layer.  Finally, the heat flux leaving the core is computed based on the temperature difference between $T_c$ and $T_{sob}$, although we use the Rayleigh number and thickness of the solid layer. The Rayleigh numbers for each layer involve their viscosities, and we take the values that we compute in section 2.2. If the tidal heating in a given layer is such that its adiabatic temperature profile is brought above the temperature of the layer below, the flux between the layers is set to zero and the temperature of the layer below is set equal to that above.     



\subsection{Orbital evolution}

We couple the orbital evolution with thermal evolution using the differential equations from \citet{Jackson2008}: 

\begin{multline}
\label{Eqn25}
\frac{1}{e}\frac{de}{dt} = -a^{-\frac{13}{2}}
\Biggl[ \frac{63}{4}\left( GM_{s}^{3}\right)^{\frac{1}{2}}\frac{R_{p}^{5}}{Q_{p}M_{p}} \\
+ \frac{171}{16} \left( \frac{G}{M_{s}}\right)^{\frac{1}{2}} \frac{R_{s}^{5}M_{P}}{Q_{s}} \Biggr]
\end{multline}

\begin{multline}
\label{Eqn26}
\frac{1}{a}\frac{da}{dt} = -a^{-\frac{13}{2}}
\Biggl[ \frac{63}{2}\left( GM_{s}^{3}\right)^{\frac{1}{2}}\frac{R_{p}^{5}}{Q_{p}M_{p}} e^{2} \\
+ \frac{9}{2} \left( \frac{G}{M_{s}}\right)^{\frac{1}{2}} \frac{R_{s}^{5}M_{P}}{Q_{s}} \Biggr]
\end{multline}

\noindent In the above equations, $Q_{p}$ and $Q_{s}$ represent tidal dissipation factors for the planet and the star, respectively. The planetary mass and radius are given by $M_{p}$ and $R_{p}$, while the stellar counterparts are represented by $M_{s}$ and $R_{s}$. The terms involving $Q_{p}$ represent tides raised on the planet, while those involving $Q_{s}$ are for tides raised on the stellar surface. We use a value of $Q_{s} = 10^{7}$ for our simulations, which is an average dissipation factor we can expect for F, G and K stars \citep{Penev2012, Lee2017}. 

We assume that the model planets form near the inner edge of the protoplanetary disk, or migrated there via disk migration. In \citet{Lee2017} it is stated that there can be a distribution of values for the inner edge of the protoplanetary disk, down to a period of one day. Therefore, we use a spread of starting points from 0.01 A.U. to 0.1 A.U. If the planets are perturbed, their orbits can become eccentric. Following \citet{Schmidt2024}, we introduce a minimum eccentricity $e_{\rm min}$ which we enforce to represent eccentricity forcing from companion planets in the system. A minimum eccentricity ensures that the orbital eccentricity does not drop below a specified value during migration. We use $e_{\rm min}$ values of $10^{-1}$, $10^{-2}$, $10^{-3}$, and $10^{-4}$. For eccentricities above 0.15, it is likley the planets will in a state of pseudo-synchronous rotation. At small $e_{min}$ values, libration is likely. Hence, our 1D thermal evolution model does not depend upon the assumption of a synchronous rotation.

\subsection{Simulation grid}

We model migration scenarios based on a starting point where the planetary mantle is fully molten and the planet has been perturbed into an eccentric orbit. We carry out simulations over a grid of initial semi-major axes ($a_{0}$) from 0.01 to 0.1 AU and initial eccentricities ($e_{0}$) from 0.01 to 0.99. We assume that, at their initial $a_{0}$, the orbital periods are shorter than the stellar rotational periods so that outward migration does not occur \citep{Lee2017}. The initial $Q_{p}$ of the mantle is computed at the start of the simulation. The tidal dissipation in the first time step of the simulation is calculated and added to the mantle energy budget. Additionally, the $Q_{p}$ is used in equations \ref{Eqn25} and \ref{Eqn26} to compute $\frac{de}{dt}$ and $\frac{da}{dt}$. Once an iteration is complete, the mantle energy budget is updated. 

The mantle temperature and orbit evolution are integrated simultaneously with an adaptive time step that is the same for all time-dependent equations. The time step for thermal evolution is limited by the largest heat flux, with a 5 percent tolerance level for the maximum temperature variation. The time step for orbital evolution is limited by the largest change in $e$ with a tolerance level of 5 percent for $\frac{\Delta e}{de/dt}$. We take the smallest time step between these to ensure for stable evolutions. The temperatures are advanced in time using a simple Euler explicit scheme with a small Courant number. 

The iterative process is continued for 10 Gyrs of simulated time. We aim to identify the initial conditions ($a_{0}$, $e_{0}$) that result in a migration history that places a lava planet at its currently observed orbital radius. The conditions are as follows:  

\begin{enumerate}
    \item The simulated planet must reach its current observed location $a_{\mathrm{final}}$.
    \item If the orbit crosses the Roche radius of the host star within the stellar lifespan, the simulation is terminated because that planet would not have survived.
    \item The timescale of migration to reach $a_{\mathrm{final}}$ cannot be greater than the measured age of the host star.
    \item If the simulation has not violated criterion 2 within the stellar lifespan, and has reached $a_{\mathrm{final}}$, the planet must not pass through the current orbit so quickly that the orbital decay rate $\delta P$ would be detectable (we set this decay rate limit at $\delta P > 10^{-3}$ seconds per year). 
\end{enumerate}

\noindent For criterion 1 it should be noted that if the planet reaches $a_{\mathrm{final}}$ in less time than the measured stellar age, we consider it a success. Even though we stop the simulation once the planet reaches $a_{\mathrm{final}}$, the orbit will continue to decay until it collapses into the star. But this would take longer than the measured stellar ages. We apply our model to seven known lava planets given in Table \ref{tab2}.

\section{Results} \label{sec3}

\begin{figure*}\centering
\includegraphics[width=1.0\textwidth]{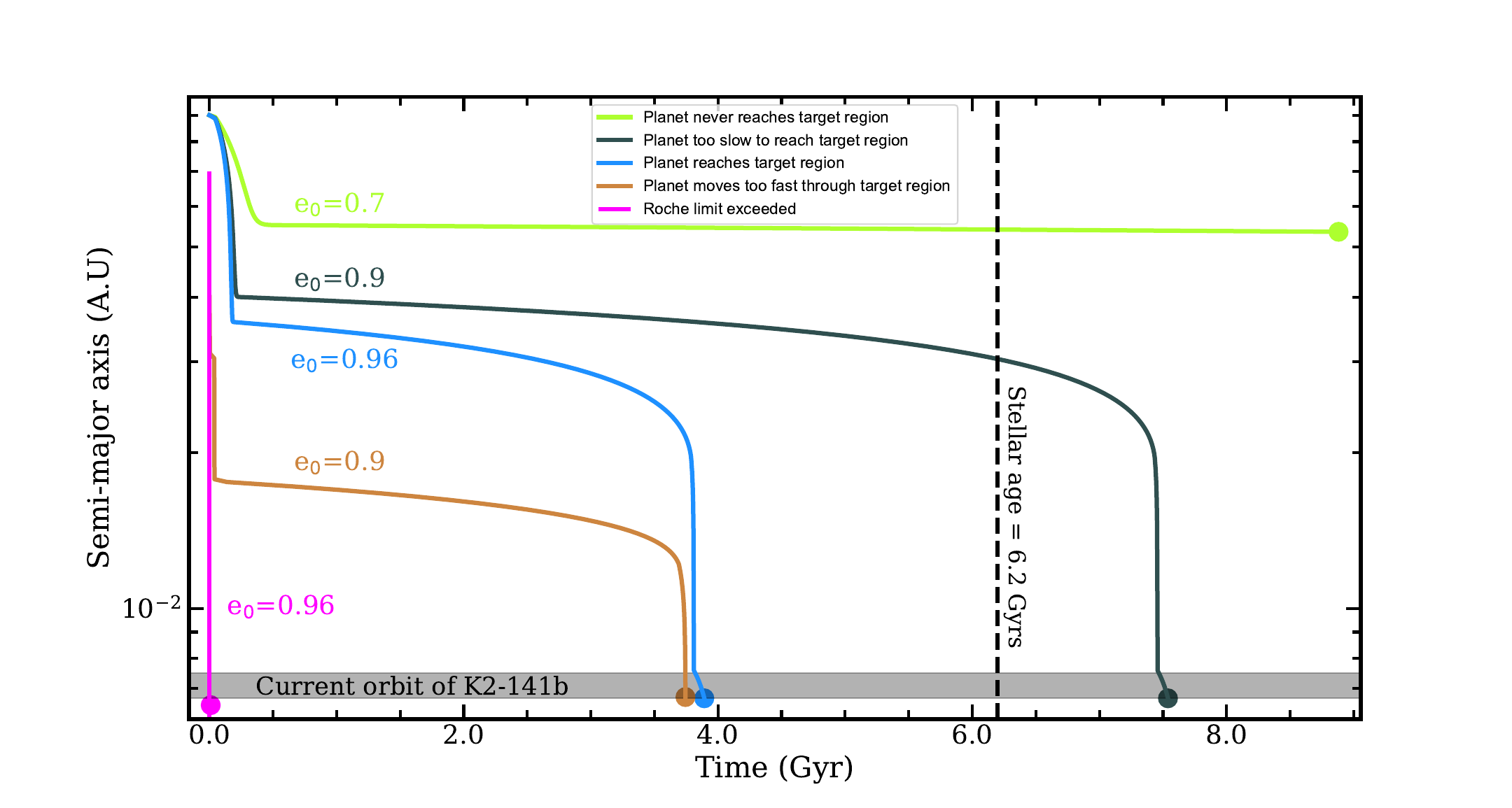}
\caption{Migration scenarios for K2-141b starting from semi-major axes of 0.09 and 0.04 AU and initial eccentricities between 0.7 and 0.96. The plots show tests at $e_{\mathrm{min}} = 10^{-2}$. Each coloured line represents a different final outcome for the simulation. The light green line} shows a simulation where the planet did not reach the present orbit of the planet after 10 Gyrs. The dark green line shows a planet that reached the observed region, but took longer than the measured age of the host star. The sky blue line shows a successful migration scenario. The dark orange line shows a planet which moved too quickly through the observed orbit: such a fast migration would produce detectable orbital decay on timescales of years. The magenta line shows a planet that entered the star's Roche radius, and got destroyed, during migration. The colours used in this figure will be re-used in Figures \ref{fig4} and \ref{fig5}.
\label{fig3}
\end{figure*}

We find in our simulations that for migration to occur from any initial semi-major axis greater than 0.01 AU, a nonzero minimum eccentricity is required. For K2-141b, simulations with $e_{\mathrm{min}} = 10^{-2}$ leads to different outcomes depending on the choices of $a_{0}$ and $e_{0}$. These outcomes can be summarized in terms of the five migration scenarios shown in Figure \ref{fig3}. Very high initial eccentricities ($e_{0} \ge 0.9$) are required to migrate the planet from $a_{0} \approx 0.1$ A.U. to the planet’s current observed orbit within K2-141’s age. Simulations with identical $e_{0}$ and $e_{\mathrm{min}}$ but $a_{0} \approx 0.05$ A.U. lead to overly rapid migration through the presently observed orbit. 

\begin{figure}\centering
\includegraphics[width=\columnwidth]{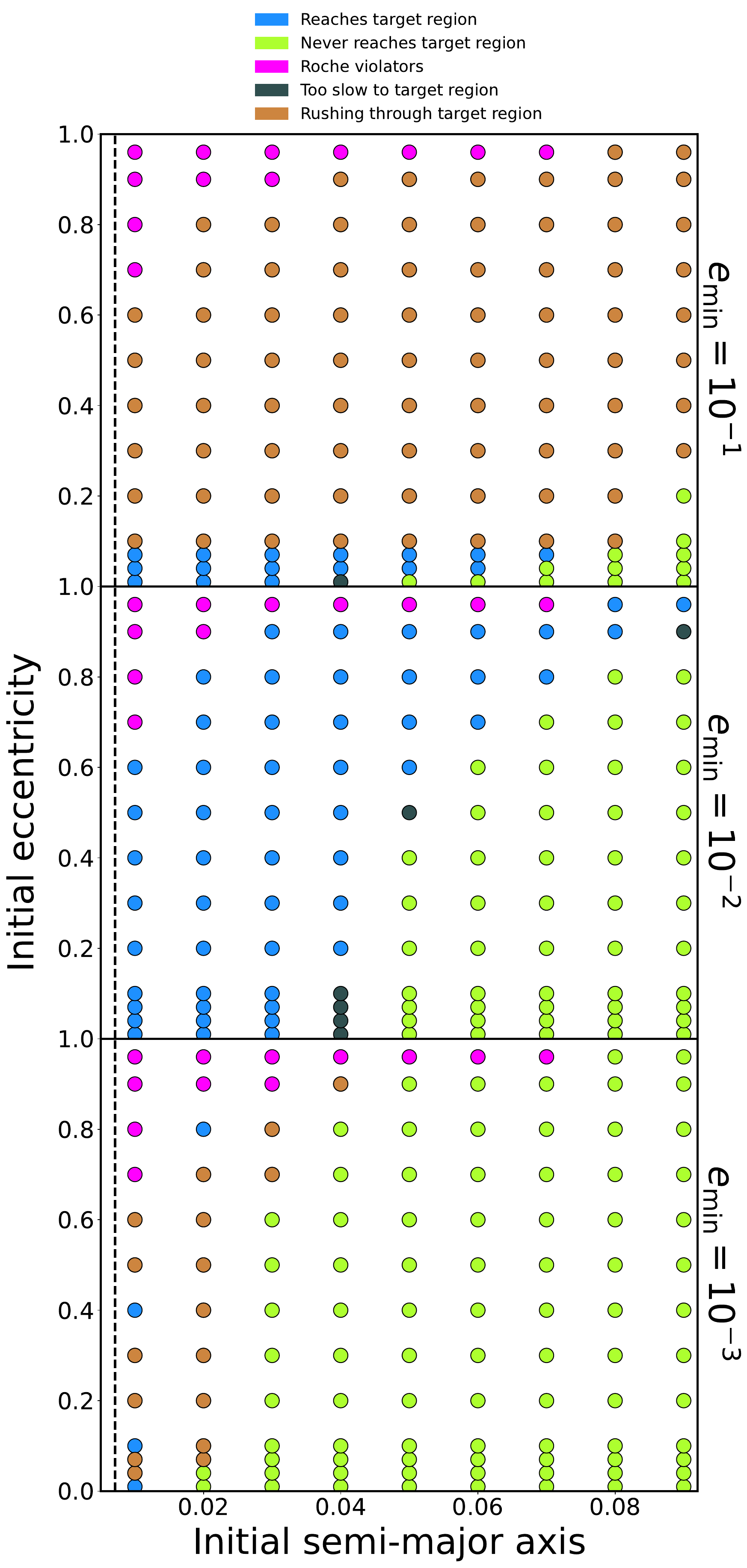}

\caption{Outcomes of orbital migration simulations for K2-141b for a variety of initial orbital parameters and minimum eccentricities of $10^{-1}$, $10^{-2}$ and $10^{-3}$. The colours match those used in Figure \ref{fig3}. The black broken vertical line on the left of the plots denotes the current observed location of K2-141b. The results show that K2-141b could only have migrated to its current orbit from the inner edge of the protoplanetary disk if it acquired an eccentricity greater than or equal to 0.9, and if the eccentricity was maintained at $10^{-2}$ for several Gyrs.}
\label{fig4}
\end{figure}

From Figure \ref{fig4} we see that $e_{\mathrm{min}} = 10^{-1}$ leads to fast migration rates, while $e_{\mathrm{min}} = 10^{-3}$ leads to extremely slow migration rates. Thus, the optimal minimum eccentricity for K2-141b is around $10^{-2}$ (although it is different for other planets we simulated). {\citet{Barragan2018} proposed a present-day eccentricity median for K2-141b of 0.06 and a 99 percent confidence upper limit of 0.20. However, observations are consistent with a circular orbit or an eccentricity too small to detect \citep{Barragan2018, Bonomo2023}.} If K2-141b experienced eccentricity forcing via a resonant orbit with K2-141c, there is no observational evidence at the moment. Furthermore in Figure \ref{fig4}, the column for $a_{0} = 0.01$ AU in the panel for $e_{\mathrm{min}} = 10^{-3}$ shows sharp changes and discontinuities between migration scenarios. As $e_{0}$ is reduced from high to low values, migration outcomes alternate between rushing through the current orbit (orange) and reaching $a_{\mathrm{final}}$ within the age of the star (blue). This non-linear behaviour is due to the feedback between interior structure and migration. The more solid and mush there is in the mantle, the higher the rate of migration. High rates of migration lead to more tidal dissipation, therefore create more melt in the mantle. The increased quantity of liquid in turn slows down the rate of migration.

\begin{figure*}\centering
\includegraphics[width=1.0\textwidth]{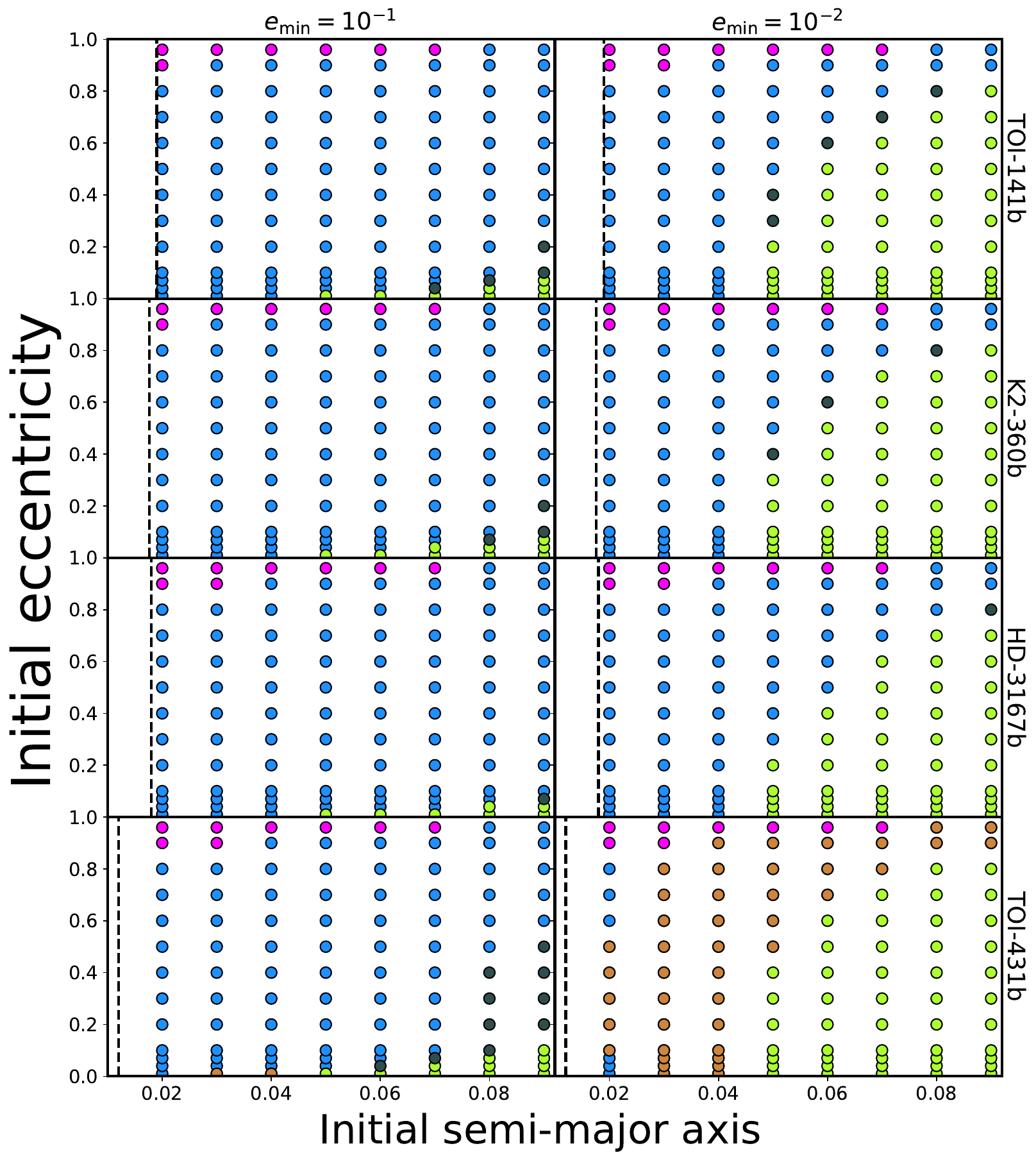}

\caption{Outcomes of orbital migration simulations for planets TOI-141b (first row), K2-360b (second row), HD-3167b (third row) and TOI-431b (fourth row), for $e_{\mathrm{min}} = 10^{-1}$ (left column) and $e_{\mathrm{min}} = 10^{-2}$ (right column). The black broken vertical line denotes the current observed location for each planet. The results show that for $e_{0} \ge 0.9$, with $e_{\rm min} \ge 10^{-2}$, we obtain successful migration scenarios for most, but not all, known lava planets. If $e_{\rm min} < 10^{-2}$, the planets do not migrate successfully to their current orbits.}
\label{fig5}
\end{figure*}

Figure \ref{fig5} presents the search grids with $e_{\mathrm{min}}$ values of $10^{-1}$ and $10^{-2}$ for TOI-141b, HD-3167b, K2-360b, and TOI-431b. For $e_{\rm min} = 10^{-2}$, most planets show trends similar to K2-141b, with the exception of TOI-431b and GJ 367b. However, for $e_{\mathrm{min}} = 10^{-1}$, several planets exhibit successful migrations, unlike K2-141b. TOI-141b, HD-3167b, and K2-360b display more successful outcomes for high $a_{0}$ and $e_{0}$ values than K2-141b. TOI-2431b follows the general trend of all lava planets except TOI-431b and GJ 367b. TOI-431b requires a particularly high $e_{\mathrm{min}}$ to reach its present orbit, and GJ-367b shows fewer successful migrations for both $e_{\rm min}$ values. Stellar age also influences migration: K2-141, TOI-141, HD\,3167, K2-360, and GJ\,367 are all older than 3 Gyr, while TOI-2431 lies between 2–10 Gyr. TOI-431 on the other hand is less than 3 Gyrs. Older systems thus favour successful migration scenarios for lava planets  through prolonged tidal and dynamical interactions.

We conducted additional simulations keeping the starting conditions the same, but reducing the tidal quality factor of the star from $10^{7}$ to $10^{5.5}$ \citep{Jackson2008, Penev2012} and removing the eccentricity minimum. In this instance, small starting eccentricities ($e_{0} < 0.0001$) result in successful migrations for $0.01 < a_{0} < 0.03$. These migrations are driven primarily through tides raised on the star. However, the magnitude of $Q_{s}$ can increase as the star gets older, which was not taken into account in this experiment. This would reduce the effect of stellar tides unless the planet gets very close to the star.

\subsection{Evolution of the planetary interior with the orbit}

Our simulations show that the material state of the mantle affects migration rate.  Tidal heating is most efficient when the timescale of tidal deformation approaches the Maxwell relaxation timescale $\tau_m$ \citep{Desro2014, organowski20}. Hence, tidal dissipation is low for a liquid layer but can become large in mush and solid layers.  To give typical numbers, the tidal quality factor $Q$ of a liquid mantle at the orbital frequencies of our simulations is very large, $Q > 10^{12}$, while that of a solid-mush combination is in the range of $1000 < Q < 10$. The inverse dependency on $Q$ in Equations \ref{Eqn25} and \ref{Eqn26} results in slow migration for primarily liquid mantles, while solid-mush combinations lead to faster migration. 

\begin{figure*}\centering
\includegraphics[width=0.8\textwidth]{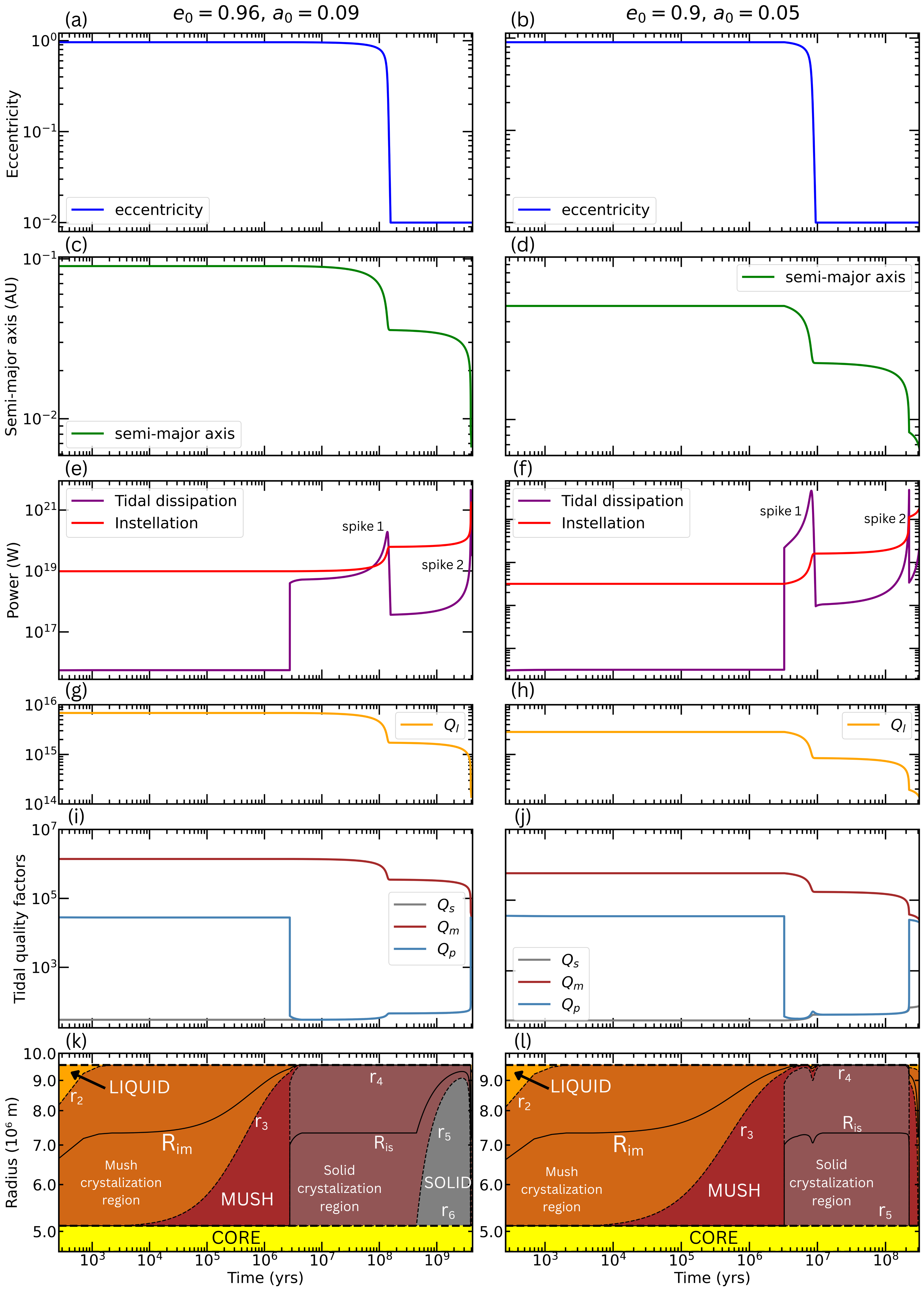}

\caption{Orbital and interior evolution of K2-141b for a minimum eccentricity of $e_{\mathrm{min}} = 10^{-2}$ and two different sets of initial conditions that leads to a successful migration of the planet to its current orbit: $a_0=0.09$ AU and $e_0=0.96$ (left column), $a_0=0.05$ AU and $e_0=0.9$ (right column). We show the time evolution of: $e$ (panels a, b); $a$ (panels c, d); tidal dissipation and instellation (panels e, f); tidal quality factors (panels g, h, i, j); and mantle cross sections (panels k, l). In panels k and l, the liquid, mush, and solid layer colours correspond with the colours in Figure \ref{fig1}. Points marked "spike 1" and "spike 2" denote periods of increased tidal dissipation. In the last row, we show the interfaces (solid black lines) from liquid to mush ($R_{\mathrm{im}}$), and mush to solid ($R_{\mathrm{is}}$). The broken black lines show the upper and lower boundary layers for the liquid-mush interface ($r_{2}, r_{3}$) and the mush-solid interface ($r_{4}, r_{5}$). The different crystallization regions are denoted in lighter shades of brown and grey to point out these transition interfaces. See \citet{Herath2024} for more details about the model.}
\label{fig6}
\end{figure*}

Figure \ref{fig6} shows the temporal evolution of $e$, $a$, $Q$, $q_{t}$, and K2-141b’s internal structure at $e_{\mathrm{min}} = 10^{-2}$ for two simulations. Both reach $a_{\mathrm{final}}$, but on different timescales. In simulation A (left column), most of the mantle solidifies after $10^{9}$ years. In simulation B (right column) a mush and liquid mantle is maintained throughout migration, with crystals of solid rock dominating the mantle for about 200 million years without fully solidifying. The rapid reduction in eccentricity (panels a and b) coincides with the first rapid drop in semi-major axis (panels c and d), representing the high-eccentricity migration stage. The subsequent epoch is the low-eccentricity migration stage, eventually leading to a second, steeper drop in semi-major axis. Spikes in tidal dissipation (marked as "spike 1" and "spike 2" in panels e and f) corresponds with drops in $Q_{p}$ (panels i and j). They mark two mantle-melting episodes for simulation B, visible in panel l, one about 10 million years in and the other taking place during the in-spiral stage after 200 million years. Including the fully liquid mantle initial condition, we have three molten periods during migration. In simulation A, we see two melting periods towards the end of migration, though it is not visible in panel k as they take place within 200 million years between 3.8 and 4 Gyrs. The sudden slowing of orbital decay in simulation B (lower right corner of panel d) occurs concurrently with the period of total mantle melting (final stages of panel l). At this stage, $Q_{p} \approx Q_{l}$, and the migration rate is significantly reduced. The same happens at the very last stages of simulation A, where $Q_{p}$ spikes upwards. This shows how the thermal state of the mantle feeds back on the orbital migration. 

The tidal heating of all simulated planets range from $10^{18}$ to $10^{24}$ W across all $e_{\mathrm{min}}$, similar to tidal heating values found in \citet{Herath2024}. As a comparison, this is significantly higher than the tidal heating in Io, of the order of $10^{14}$ W. The much larger tidal heating in lava planets reflects the combination of their larger size, their smaller orbital distance and more massive primary body. At the maximum of these heating rates, magma oceans can be deeper than 2000 km. When tidal heating is negligible, the magma oceans can still be as deep as 200–300 km as a result of the instellation energy \citep{Kite2016, Yang2025}. We observe this change in magma ocean depth in our simulations; the magma ocean rapidly shrinks during the early stages of migration, when tidal heating is negligible, but grows deep ($>$ 1000 km) towards the end due to the accelerating migration and increased tidal heating.

\begin{figure}\centering
\includegraphics[width=1.0\columnwidth]{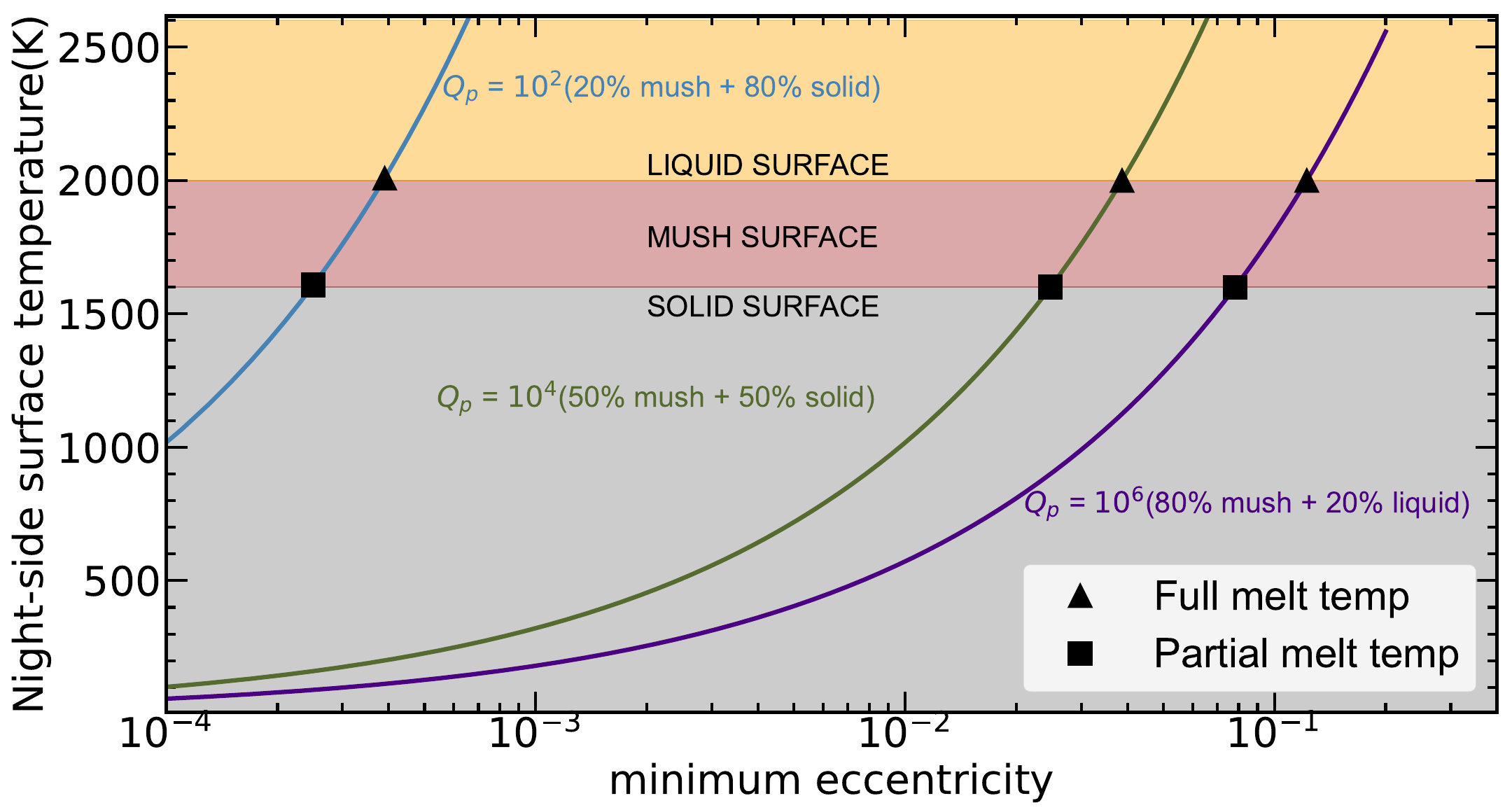}

\caption{The night-side surface temperatures for different minimum eccentricities for K2-141b, at its present day orbit. The three lines in blue, green and indigo represent three types of mantle mixtures where the global quality factor $Q_{p}$ for each was calculated following the method outlined in Section 2.1. The x-axis gives a spread of possible minimum eccentricities at the current orbit. The y-axis gives the resulting night-side surface temperatures if tidal dissipation is the only source of heat. The compositions are the same mixtures as those used in Figure \ref{fig2}. The grey, brown and orange horizontal layers show the different states of the mantle at the night-side surface (respectively, solid, mush and liquid) as a function of its temperature.}
\label{fig7}
\end{figure}

\section{Discussion and conclusions} \label{sec4}

Applying our thermal–orbital evolution model to seven known ultra-short-period planets reveals the conditions under which rocky planets become lava planets. Our model shows that if the inner edge of the protoplanetary disk is around 0.1 A.U., rocky planets cannot migrate into tight orbits without early high-eccentricity perturbations and sustained eccentricity forcing. The eccentricity forcing ($e_{\mathrm{min}} \ge 10^{-2}$) has to be continuous until the planet is within 0.01 AU of the star. If the inner edge of the protoplanetary disk is closer to the star, then eccentricity forcing may not be necessary. We find with our simulations that with a low stellar $Q$ migration is possible with a small $e_{0}$, and without eccentricity forcing. The stellar $Q$ is expected to increase by a few orders of magnitude as stars get older and their rotation rates slow down. We did not take stellar rotation rates into account when modelling the tidal forces.    

It is easier to explain a lava planet around an old star because the low-eccentricity stage of migration is slow. Our simulations have higher numbers of successful migrations within the parameter space we tested for stars older than 3 Gyrs. Older stars are more likely to have slower spin rates than the orbital periods of the migrating planets. If the planets start their journey around 0.1 A.U., the host star would be too far to feel significant tidal effects. By the time the planet gets close enough to be affected by stellar tides, the stellar spin rate would be reduced.  

Migration happens in three stages for these planets. There is disk migration to the inner edge of the protoplanetary disk, followed by two stages that we simulate: a high-eccentricity phase driven by early orbital perturbations, reducing the semi-major axis by about a factor of two, followed by a low-eccentricity stage during which the semi-major axis shrinks by a factor of five. For six of seven modelled planets, our results show that this two-stage process, governed by tidal dissipation and eccentricity forcing, is a viable migration pathway for lava planets. The exception, TOI-431b, moves through its current orbit too rapidly for most initial $a_{0}$ and $e_{0}$ values. Only when $e_{\mathrm{min}} = 10^{-1}$ did TOI-431b exhibit successful migration by slowing down as it approached its current orbit. This is because the higher minimum eccentricity led to increased tidal dissipation and therefore more mantle melt. A molten mantle slows the migration. This behavior reflects the coupling between orbital mechanics and internal structure evolution.  

Planetary tidal dissipation, the cause of migration in our scenarios, requires a non-zero eccentricity. A non-zero eccentricity, in turn, must be maintained by gravitational interactions with other planets. Hence, our simulations suggest that gravitational interactions between lava planets and companion planets are key to their migration. The planets considered in this study all have more massive companion planets with orbital periods of 4 to 9 days \citep{Barragan2018, Malavolta2018, Espinoza2019, Livingston2018}, which likely shaped their dynamical evolution. A central question is whether perturbations from such companions could generate the high early eccentricities required by our scenarios. \citet{Petro2018} demonstrated that secular chaos can produce initial eccentricities up to $e = 0.9$, and simulations with companions at period $P < 10$ days can push low-mass inner planets into ultra-short-period orbits. \citet{Livingston2024} further showed that high-eccentricity migration and obliquity tides \citep{Mill2020} are viable formation pathways for K2-360b, with obliquity tides favoured due to an inclined, twice-as-massive outer companion at $P < 10$ days. 

It is possible for lava planets to go off-resonance from their companion planets as the former gets closer to the star \citep{Schmidt2024}, removing the minimum eccentricity. We can deduce if the minimum eccentricity remains today by observing the night-side temperature of these planets. In Figure \ref{fig7} we show possible temperatures for the night-side surface of K2-141b at its current orbit, assuming that tidal dissipation is the only source of heating. The dissipation in this case is powered by the minimum eccentricity. We use Equation \ref{Eqn16}, where for $Q_{p}$ we include the quality factors obtained for the three types of mantles of K2-141b with different mixtures, as done in Section 2.1 and Figure \ref{fig2}. Since these are global quality factors, we use the radius of K2-141b for $R_{p}$. For $e$ we use a spread of possible minimum eccentricities from $10^{-4}$ to $10^{-1}$ at its current orbit. The figure shows the temperatures required to melt the night-side surface for a thin magma or mush ocean. If the current eccentricity is greater than $10^{-4}$, we can expect a sizeable night-side surface temperatures. A renewed exploration of the dynamical histories of these systems with the inclusion of varying tidal quality factors and tidal heating will shed more light on the mechanisms that drive rocky planets into close, short-period orbits.

Although we have shown that our coupled thermal-orbital evolution model can explain the inward migration of lava planets, some of its limitations must be acknowledged. First, our migration scenarios require an ever present minimum eccentricity, which we impose instead of letting it arise naturally from gravitational interactions with one or more companions. Second, we used fixed values for the viscosity of the solid, mush and liquid states.  More realistically, the viscosity of each mantle layer should also be temperature dependent, and the viscosity of the mush should depend on its local melt fraction, which can vary with radius. Similarly, we assume a silicate density and other thermodynamical parameters that are representative, but fixed; a more realistic model would allow for these to depend on  both the phase and pressure.  The dynamical effects caused by density contrasts between solid and liquid phases was explored in \citet{Boukare2025b}, and should ideally be included in thermal-orbital evolution models. Finally, we assume that migration begins immediately after planetary formation, rather than allow for the planet to cool down for a few billion years. All, or some, of these limitations can hopefully be tackled in future thermal-orbital evolution models.

\begin{acknowledgments}

We would like to thank Professor Eve J. Lee at the University of California, San Diego for her valuable insights and input to this paper on planet migration theories. N.B.C. acknowledges support from an NSERC Discovery Grant, a Tier 2 Canada Research Chair, and an Arthur B. McDonald Fellowship. M.H would like to thank the Fonds de recherche du Qu\'ebec for a doctoral fellowship.  M.D. acknowledges support from an NSERC Discovery Grant. The authors also thank the Trottier Space Institute and l'Institut Trottier de recherche sur les exoplan\'etes for their financial support and dynamic intellectual environment.

\end{acknowledgments}

\begin{contribution}

M.H. constructed the computational model and ran simulations for the interior structure and orbital mechanics, and wrote the manuscript. N.B.C. conceived the project, designed the research plan, guided the direction of the project, and contributed to the manuscript. C-\'E.B. designed the mathematical framework for the thermal evolution model and provided guidance on computational aspects. M.D. provided the mathematical framework for the Andrade model and contributed to the manuscript.

\end{contribution}

\appendix \label{ap}

\renewcommand{\thetable}{A\arabic{table}}

\begin{table*}\centering
  \begin{tabular}{c c c c c}
   \hline
    Parameter & Description & Value & Units & Reference  \\ 
     \hline
    \\
  
    $\mathrm{C_{core}}$ & specific heat capacity of Fe core & 850 & J $\mathrm{Kg}^{-1} \mathrm{K}^{-1}$ & \cite{Sc2016}   \\
    $\mathrm{C}$ & specific heat capacity of silicates & 1260 & J $\mathrm{Kg}^{-1} \mathrm{K}^{-1}$ & \cite{Sc2016}   \\
    $\rho$ & Bulk density of silicates & 4000 & Kg $\mathrm{m}^{-3}$ & \cite{Leb2013}  \\
    $\sigma$ & Stefan-Boltzmann constant & $5.67 \times 10^{-8}$ & W $\mathrm{m}^{-2} \mathrm{K}^{-4}$ \\
    $\kappa$ & thermal conductivity of magma & 13 & W $\mathrm{m}^{-1} \mathrm{K}^{-1}$ & \cite{Ohta2017} \\ 
    $\alpha_{c}$ & thermal expansivity (core) & $2 \times 10^{-5}$ & $\mathrm{K}^{-1}$ & \cite{Herath2024} \\
    $\alpha_{m}$ & thermal expansivity (mantle) & $3 \times 10^{-5}$ & $\mathrm{K}^{-1}$ & \cite{Tach2011} \\
    $\kappa_{d}$ & thermal diffusivity & $7.5 \times 10^{-7}$ & $\mathrm{m}^{2} \mathrm{s}^{-1}$ & \cite{Sc2016} \\
    $\eta_{m}$ & viscosity of magma & 0.1 & Pa s & \cite{Boukare2023} \\
    $\delta$ & relative viscosity & 25.7 & Pa s & \cite{Kervazo2021} \\
    $\xi$ & part of auxiliary function & $1.17 \times 10^{-9}$ & - & \cite{costa2009} \\
    $\gamma$ & rapidity of relative viscosity increase & 5 & Pa s & \cite{costa2009} \\
    $\phi_{*}$ & critical melt fraction & 0.569 & - & \cite{costa2009} \\
    B & Einstein coefficient & 2.5 & - & \cite{costa2009} \\
    $\mu_{r}$ & mean shear modulus & $1.5 \times 10^{11}$ & Pa & \cite{Castillo2011} \\
    $\alpha$ & Andrade exponent & 0.33 & - & \cite{Tobie2019} \\
    \\
 
     \hline
  \end{tabular}

  \caption{Table of constants.}
  \label{tab1}
   
\end{table*}

\begin{table*}\centering
  \begin{tabular}{l l l l l r}
   \hline
    Planet & Mass ($\mathrm{M_{\oplus}}$) & Radius ($\mathrm{R_{\oplus}}$) & Period (days) & Stellar age (Gyrs) & Reference  \\ 
     \hline
    \\
  
    K2-141b & $5.16_{-0.35}^{+0.35}$ & $1.54_{-0.09}^{+0.10}$ & $0.2803244 \pm 0.0000015$ & $6.20_{-4.70}^{+6.60}$ & \citet{Barragan2018}  \\
    TOI-141b & $8.83_{-0.66}^{+0.65}$ & $1.750_{-0.051}^{+0.052}$ & $1.0080351_{-0.0000020}^{+0.0000021}$ & $4.00_{-0.79}^{+0.66}$ & \citet{Espinoza2019}  \\
    TOI-431b & $3.07\pm0.35$ & $1.28\pm0.04$ & $0.490047_{-0.000007}^{+0.000010}$ & $1.90\pm0.30$ & \citet{Osborn2021} \\
    TOI-2431b & $6.20 \pm 1.20$ & $1.54 \pm 0.03$ & $0.22419577\pm 0.00000001$ & $2.0_{-1.7}^{+9.1}$ & \citet{HanTas2025} \\ 
    HD-3167b & $4.97_{-0.23}^{+0.24}$ & $1.67_{-0.10}^{+0.17}$ & $0.959641 \pm 0.000011$ & $7.80 \pm 4.30$ & \citet{Dai2019} \\
    K2-360b & $7.67 \pm 0.75$ & $1.565 \pm 0.079$ & $0.87724 \pm 0.00004$ & $6.0_{-2.9}^{+2.6}$ & \citet{Livingston2024} \\
    GJ-367b & $0.63 \pm 0.05$ & $0.69 \pm 0.02$ & $0.3219225 \pm 0.0000002$ & $7.95 \pm 1.31$ & \citet{Lam2021} \\
 
    \\
 
     \hline
  \end{tabular}
  
\caption{List of planets modelled in this paper.}
\label{tab2}
   
\end{table*}


\bibliography{sample701}{}
\bibliographystyle{aasjournalv7}


\end{document}